# Normal State Resistivity of Underdoped $YBa_2Cu_3O_x$ Thin Films and $La_{2-x}Sr_xCuO_4$ Ultra-Thin Films under Epitaxial Strain

**Lieven Trappeniers, Johan Vanacken, Patrick Wagner, Gerd Teniers, Severiano Curras, Joel Perret[*], Pierro Martinoli[*], Jean-Pierre Locquet[**], Victor V. Moshchalkov and Yvan Bruynseraede**

*Laboratorium voor Vaste-Stoffysica en Magnetisme,*
*Katholieke Universiteit Leuven, Celestijnenlaan 200 D, B-3001 Leuven*
[*]*Institut de Physique, Université de Neuchâtel, CH-2000 Neuchâtel, Switzerland*
[**]*IBM Research Division, Zürich Research Laboratory, CH-8803 Rüschlikon, Switzerland*

*The normal state resistivity of high temperature superconductors can be probed in the region below $T_c$ by suppressing the superconducting state in high magnetic fields. Here we present the normal state properties of $YBa_2Cu_3O_x$ thin films in the underdoped regime and the normal state resistance of $La_{2-x}Sr_xCuO_4$ thin films under epitaxial strain, measured below $T_c$ by applying pulsed fields up to 60 T. We interpret these data in terms of the recently proposed 1D quantum transport model with the 1D paths corresponding to the charge stripes.*

## 1. INTRODUCTION

In order to understand why superconductivity arises in the high temperature superconductors (HTS), one first needs to know the underlying normal state properties of these materials below their critical temperature $T_c$. The normal as well as the superconducting state are strongly dependent on doping and thus the carrier concentration.

In this work, two methods for changing $T_c$ will be applied. In the $YBa_2Cu_3O_x$ system, we will change the oxygen content whereas in the $La_{2-x}Sr_xCuO_4$ system we will tune $T_c$ by changing the lattice mismatch of the $La_{1.9}Sr_{0.1}CuO_4$ films with the substrate [1] at fixed Sr content $x = 0.1$. High field transport measurements were performed using the pulsed magnetic fields technique [2].

## 2. NORMAL STATE TRANSPORT IN $YBa_2Cu_3O_x$ THIN FILMS

While studying a series of underdoped $YBa_2Cu_3O_x$ thin films, a remarkable scaling of the zero field resistivity on a universal curve was observed (figure 1). Although reported previously [3], we have now identified three distinct regions in this universal behaviour (figure 1). *The simple linear behaviour in region I* was already noted from the first measurements on optimally doped HTS. *The super-linear behaviour in region II*, we will attribute to the opening of a spin gap [4]. This behaviour, is barely present in the optimally doped samples, but it becomes very pronounced in underdoped HTS samples. The third dependency which we observe, only visible in the underdoped samples at low temperatures, is a *logarithmically diverging resistance with lowering temperature*.

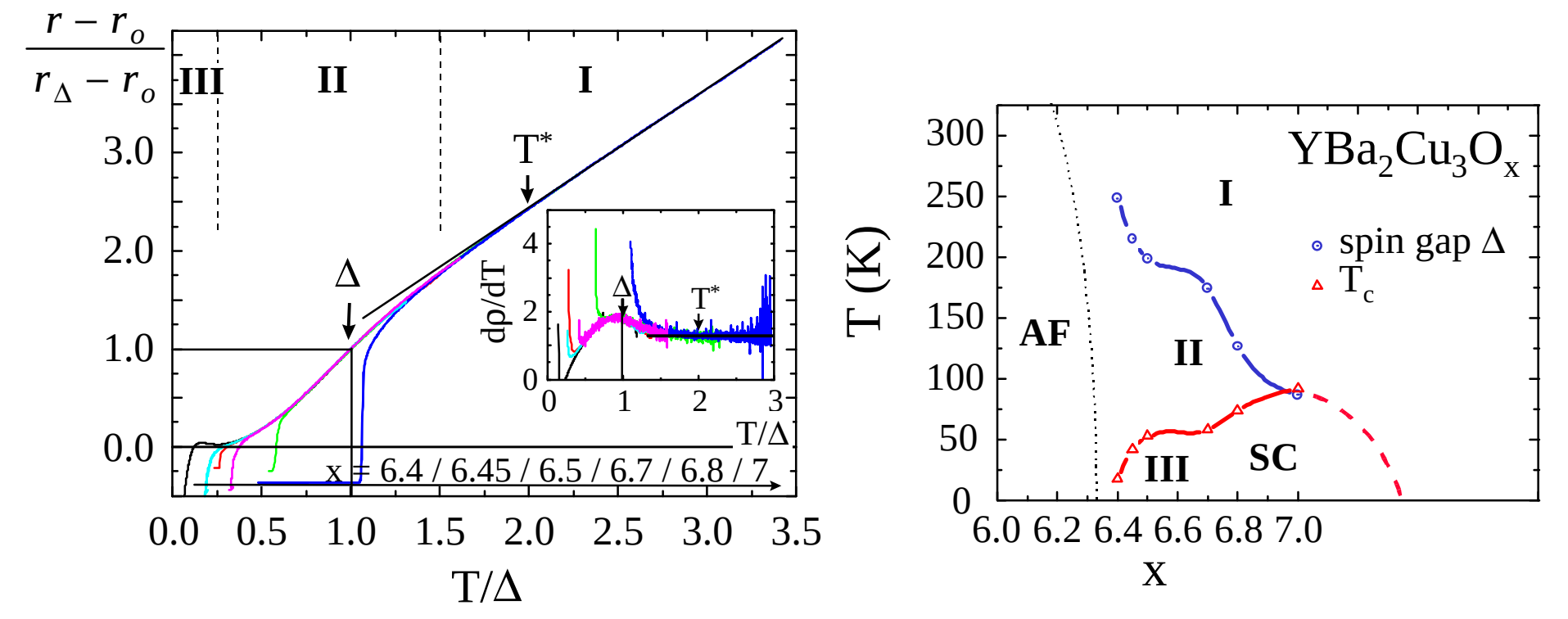

| Region III | Region II | Region I |
|---|---|---|
| $r(T) = r_0 + r_1 \ln\left(\frac{1}{T}\right)$ | $r(T) = r_0 + C\frac{T}{\Delta}\exp\left(-\frac{\Delta}{T}\right)$ | $r_{2D}^{-1}(T) = s_{2D}(T)\Big|_{L_f = x_m} \propto \frac{J}{T}$ |

Fig. 1. (left plot) Universal *r(T)* behaviour in underdoped $YBa_2Cu_3O_x$; (right plot) Critical temperature $T_c$ and spin gap *D* versus the oxygen content *x* of $YBa_2Cu_3O_x$. The proper *r(T)* expressions [4,5] for the 3 regions are given; *D* is the spin gap and *J* indicates the interaction energy.

More features of the model used to describe the measurements [4,5] are summarised in figure 1. The model is based on the following basic principles: The Cu-O HTS, all have $CuO_2$ planes as building blocks. Such an undoped $CuO_2$ plane is a 2D antiferromagnet. When holes are introduced in the $CuO_2$ plane, the AF ground state will be perturbed. In the case of low doping, it will be very difficult for the holes to become mobile, and they will be localised, surrounded by an AF background. With larger amounts of holes, a phase segregation between charge areas and AF areas will occur. It is however clear that charge transport must be strongly influenced by magnetic scattering. Therefore, it has been proposed in Ref. [4] that (i) the dominant scattering mechanism is of magnetic origin. (ii) The resistivity is determined by the inverse quantum conductivity $s^{-1}$ with (iii) the inelastic scattering length $L_\phi$ controlled by the magnetic correlation length $x_m$. The expression of the magnetic correlation length $x_m$ is dependent on the effective dimensionality of the system. The three distinct regions in figure 1 can then be associated with a 2D Heisenberg regime (I), a 1D stripe phase (II) and a localised (2D) regime (III), respectively.

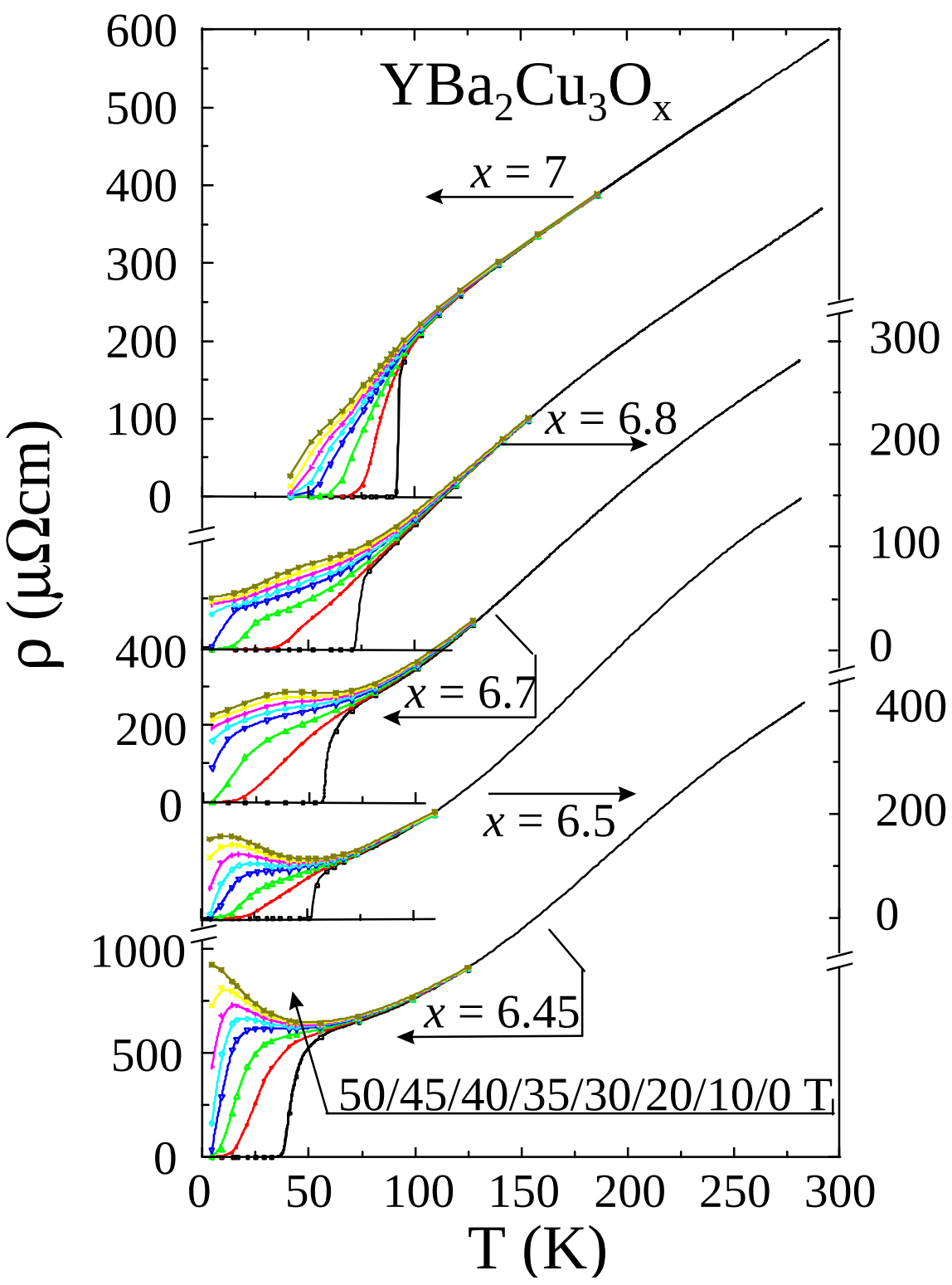

Fig. 2. *r(T)* at fixed fields for the underdoped $YBa_2Cu_3O_x$ samples.

From the scaled *r(T)* curves, the real values of the spin gap *D* were found by applying the proper equation in region II (see the phase diagram in figure 1). From this plot it is clear that the superconducting state masks the underlying transport properties of the HTS. Therefore, *r(H)* experiments in pulsed magnetic fields were performed and from these the *r(T)* behaviour at fixed field were extracted, as shown in figure 2.

In figure 2 it can be observed that at high doping levels, the $YBa_2Cu_3O_x$ thin films show a metallic behaviour in their *ab* plane, even at 50 T. When the charge carrier concentration is decreased, regions II and III (as introduced in figure 1) can be easily identified.

## 3. NORMAL STATE TRANSPORT IN $La_{1.9}Sr_{0.1}CuO_4$ THIN FILMS UNDER EPITAXIAL STRAIN

The advantage in using strained $La_{2-x}Sr_xCuO_4$ thin films, is that for a fixed stoichiometry, the lattice dimensions can be changed (enlarged or decreased), thus strongly affecting the critical temperatures $T_c$ [1]. Here we present data on three samples with $x = 0.1$. Sample A and B are respectively 125 and 150 Å thin films, grown on $SrLaAlO_4$ whereas sample C is a 150 Å thin film grown on $SrTiO_3$. Using a $SrLaAlO_4$ substrate leads to compressive strain, since the lattice parameters of $SrLaAlO_4$ are smaller than those of $La_{1.9}Sr_{0.1}CuO_4$. When growing films on $SrTiO_3$: tensile stress is induced due to the larger in-plane parameters of $SrTiO_3$. The influence on $T_c$ of the induced stress can be seen in figure 3 (left). By applying tensile stress the *ab*-lattice parameters increase and the *c*-axis lattice parameter decreases, leading to reduced $T_c$ values. By applying compressive stress, the opposite structural effects occur, leading to an increase of $T_c$.

Also in this compound the three distinct regions can qualitatively be observed in the *r(T)* behaviour, as shown in figure 3 (left). Moreover, in the right plot it is shown that, not only does the *r(T)* expression for the 1D stripes fit good, also a scaling between the $Sr_{2.5}Ca_{11.5}Cu_{24}O_{41}$ spin-ladder compound under pressure [6] and the underdoped $La_{1.9}Sr_{0.1}CuO_4$ is proven for $T/T_o < 0.6$. Since, in the spin ladder compound, it is known that the transport mechanisms are dominated by the 1D magnetic scattering, this further motivates the principles of our approach [4,5]. Fitting the 1D region (II) with the proper expression reveals a gap $D \approx 216$ K.

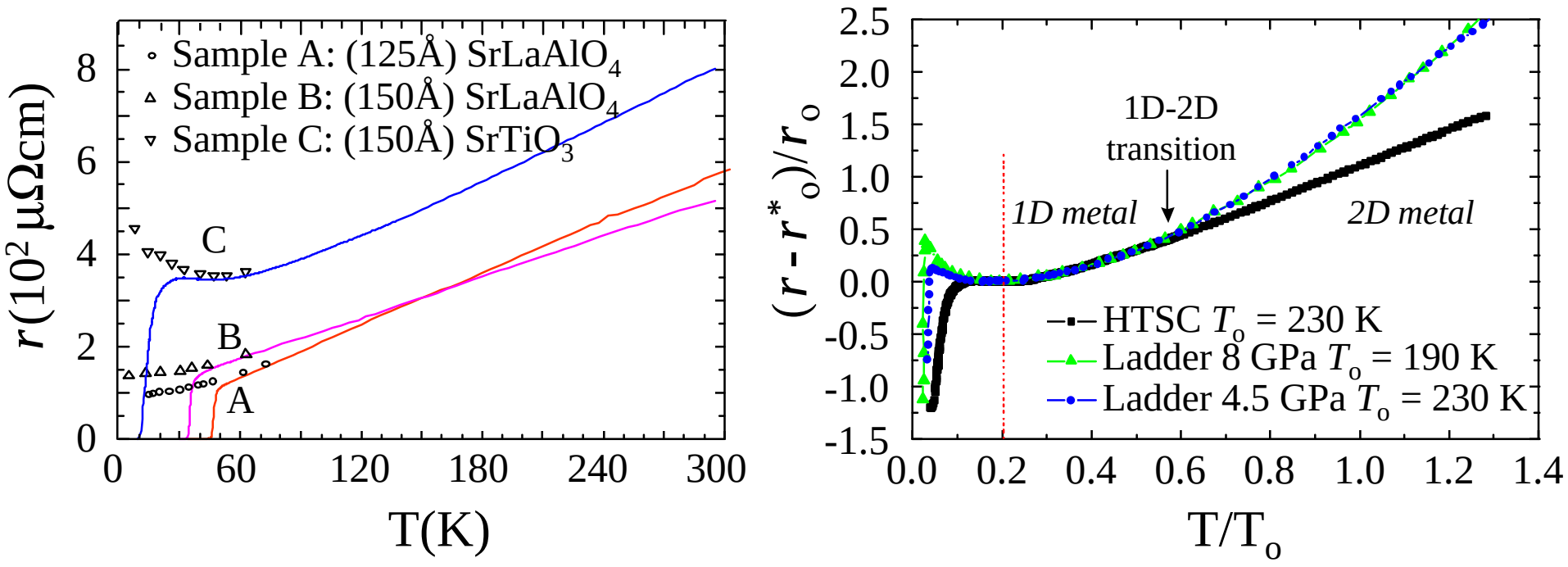

Fig. 3. (left) *r(T)* behaviour for different $La_{1.9}Sr_{0.1}CuO_4$ samples (A, B, C) and their normal state *r(T)* behaviour at $m_oH = 50$ T. (right) Relative resistivity $(r-r_o^*)/r_o$ versus $T/T_o$. The HTS and ladder compound scale with each other for $T/T_o < 0.6$.

## ACKNOWLEDGMENTS

This work is supported by the Flemish FWO, GOA, IWT and the Belgian IUAP Programs as well as the Swiss NSF project 2029-050538.97.

## REFERENCES

1. J.-P. Locquet, J. Perret, J. Fompeyrine, E. Mächler, J.W Seo and G. Van Tendeloo, *Nature* **394**, 453 (1998).
2. F. Herlach, L. Van Bockstal, R. Bogaerts, I. Deckers, G. Heremans, L. Li, G. Pitsi, J. Vanacken and A. Van Esch, *Physica B* **201**, 542 (1994).
3. B. Wuyts, V.V. Moshchalkov and Y. Bruynseraede, *Phys. Rev. B* **53**, 9417 (1996).
4. V.V. Moshchalkov, *Sol. St. Comm.* **86**, 715 (1993).; V.V. Moshchalkov, L. Trappeniers and J. Vanacken, *Europhys. Lett.* **46**, 75 (1999).; cond-mat/9902297.
5. J. Vanacken, L. Trappeniers, J. Perret, J.P. Locquet, V.V. Moshchalkov and Y. Bruynseraede, *submitted to Physica C.*
6. T. Nagata, M. Uehara, J. Goto, N. Komiya, J. Akimitsu, N. Motoyama, H. Eisaki, S. Uchida, H. Takahashi, N. Nakanishe and N. Môri, *Physica C* **282-287**, 153 (1997).